\documentclass[twocolumn,showpacs,nofootinbib,aps,10pt, pra,superscriptaddress]{revtex4-1}

\usepackage{amsmath,xcolor,amsfonts,amsthm,amssymb,graphicx, srcltx,hyperref}
\usepackage{times,todonotes}
\usepackage[normalem]{ulem} 
\usepackage{todonotes}

\def\ket #1{\vert #1\rangle}

\newcommand{\beq}{\begin{equation}}
\newcommand{\eeq}{\end{equation}}
\newcommand{\dd}{{\rm d}}
\newcommand{\PsiOut}{\ket{\Psi_{\rm out}}}

\begin{document}

\title{
Interference at the Single Photon Level Along  Satellite-Ground Channels}

\author{Giuseppe Vallone}
\author{Daniele Dequal}
\author{Marco Tomasin}
\author{Francesco Vedovato}
\author{Matteo Schiavon}
\affiliation{Dipartimento di Ingegneria dell'Informazione, Universit\`a degli Studi di Padova, Padova, Italy.}
\author{Vincenza Luceri}
\affiliation{e-GEOS spa, Matera, Italy}
\author{Giuseppe Bianco}
\affiliation{Matera Laser Ranging Observatory, Agenzia Spaziale Italiana, Matera, Italy}
\author{Paolo Villoresi}
\email{E-mail:  paolo.villoresi@dei.unipd.it}
\affiliation{Dipartimento di Ingegneria dell'Informazione, Universit\`a degli Studi di Padova, Padova, Italy.}


\begin{abstract} 
Quantum interference 
arising from superposition of states
is a striking evidence of the validity of Quantum Mechanics,
confirmed in many experiments and also exploited in applications.
However, as for any scientific theory, Quantum Mechanics is valid within the limits in which it has been experimentally verified. 
In order to extend such limits, it is necessary to observe 
quantum interference in unexplored conditions such as moving terminals at large distance in Space.
Here we experimentally demonstrate single photon interference at a ground station 
due to the coherent superposition of two temporal modes  reflected by a rapidly moving satellite thousand kilometers
  away.
The relative speed of the satellite induces 
a varying modulation in the interference pattern.
The measurement of the satellite distance in real time by laser ranging allowed us to precisely predict the instantaneous value of the 
interference phase.
We then observed the interference patterns with visibility up to 
$67\%$ with three different satellites and with path length up to 5000 km.
Our results attest the viability of photon temporal modes for fundamental tests of Physics and Quantum Communications in Space.
\end{abstract}

\maketitle

{\it Introduction -}
Quantum interference has played a crucial role to highlight the essence of Quantum Mechanics
 since the Einstein-Bohr dialogues at the end of the Twenties \cite{bohr35pr}. 
Indeed, it originates when {\it alternative possibilities} in a quantum process are indistinguishable, 
like in the case of individual particles that may be  simultaneously in more than one place as in the well-known Young double-slit
experiment~\cite{bohr35pr}.
Quantum interference
has been observed with photons~\cite{tayl09pcps,gran85prl}, but also with
electrons~\cite{jons74ajp}, neutrons~\cite{rauc15book} and even with large molecules with masses exceeding 10000 amu~\cite{eibe13pccp}. 
One of the main challenges in Quantum Physics is establishing if fundamental bounds to interference exist: for instance, 
can quantum interference  be measured  by observers in relative motion and at arbitrary large distance?
To our knowledge,
the longest path on which
interference at the single photon level was tested is a
307 km fiber link on ground with fixed sender and receiver~\cite{korz15nap}.
Classical interference in free-space has been observed in gravitational wave detectors based on a Michelson interferometer with 4 km long arms and using a laser beam with kilowatt power~\cite{Ackley15cqg}.

Here we demonstrate  { 
interference at the single photon level}
along satellite-ground channels
 by exploiting temporal modes of single photons.
To this purpose, we  exploited  a  coherent superposition between two single-photon wavepackets
on ground
and observed  their interference 
after the reflection by a rapidly moving satellite at very large distance  with a total path length up to 5000 km.
The varying relative velocity of the satellite with respect to the ground introduces
a modulation in the interference pattern  which can be predicted by special relativistic calculations, as explained below.

\begin{figure}[t]
\centering
\includegraphics[width=8cm]{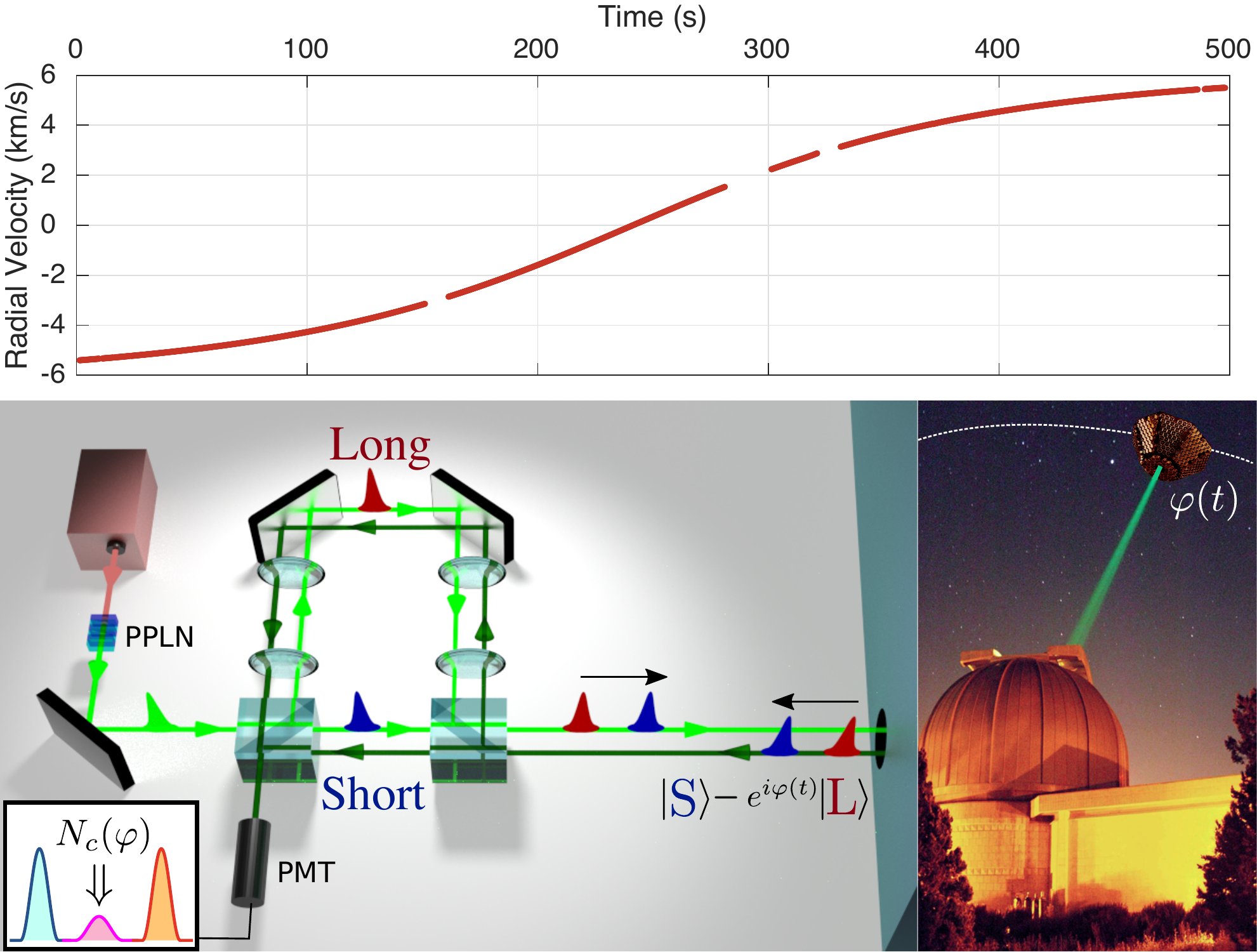}
\caption{{\bf Scheme of the experiment and satellite radial velocity}.
In the top panel we show the measured radial velocity of the Beacon-C satellite ranging from $-6$ km/s to $+6$ km/s as a function of time during a single passage.
In the bottom panel we show the unbalanced MZI with the two $4f$-systems used for the generation of the state
and the measurement of the interference.
Light and dark green lines respectively  represent the beams outgoing to and ingoing from the  telescope. 
In the inset, we show the expected detection pattern: the number of counts $N_c$ in the central peak 
varies according to the kinematic phase $\varphi$ imposed by the satellite.
Right photo shows MLRO with the laser ranging beam and the Beacon-C
satellite (not to scale).
The phase $\varphi(t)$ depends on the satellite radial velocity as described in the text.}
\label{fig:1}
\end{figure}

{\it Description of the experiment -}
In our scheme,  
 a coherent state $\PsiOut$ in two temporal modes
is generated at the ground station with an 
unbalanced Mach-Zehnder interferometer (MZI),  sketched in the lower panel of  Fig. \ref{fig:1}.
The delay $\Delta t ~\simeq3.4$ ns
between the wavepackets of the two modes
corresponds to a length difference between the two arms
 of $\ell
 = c\Delta t\simeq 1$ m  ($c$ is the speed of light in vacuum)
 and it is much longer than the 
 coherence time $\tau_c\approx83$ ps of each wavepacket
 (we used the convention of~\cite{sale91bk} for the definition of $\tau_c$).
Using a telescope, the state $\PsiOut$
is directed to a retroreflector placed on a satellite in orbit.  
 The satellite retroreflectors redirect the beam back to the ground station, where it is collected and injected into the same MZI used in the uplink. After the reflection by the satellite and the downlink attenuation, the state collected by the telescope can be written as $\ket{\Psi_r} = ({1}/\sqrt{2}) (\ket{S} - e^{i\varphi(t)}\ket{L})$, namely as a
a superposition of two single-photon wavepackets $\ket{S}$ and $\ket{L}$
(the quantum state written above corresponds to the
re-normalized single-photon part of the state received
at the telescope).
We note that the above superposition is also known as {\it time-bin encoding}, and it is
used for fundamental tests of Quantum Mechanics \cite{fran89prl,titt98prl,lima10pra,carv15prl}, for Quantum Information applications such as 
quantum key distribution (QKD) \cite{scar09rmp,bacc13nco} along optical fibers \cite{mull97apl,bren99prl,gisi02rmp}
and for increasing the dimension of the Hilbert space in which  information can be encoded~\cite{alik07prl}.

\begin{figure}[t]
\centering
\includegraphics[width=8cm]{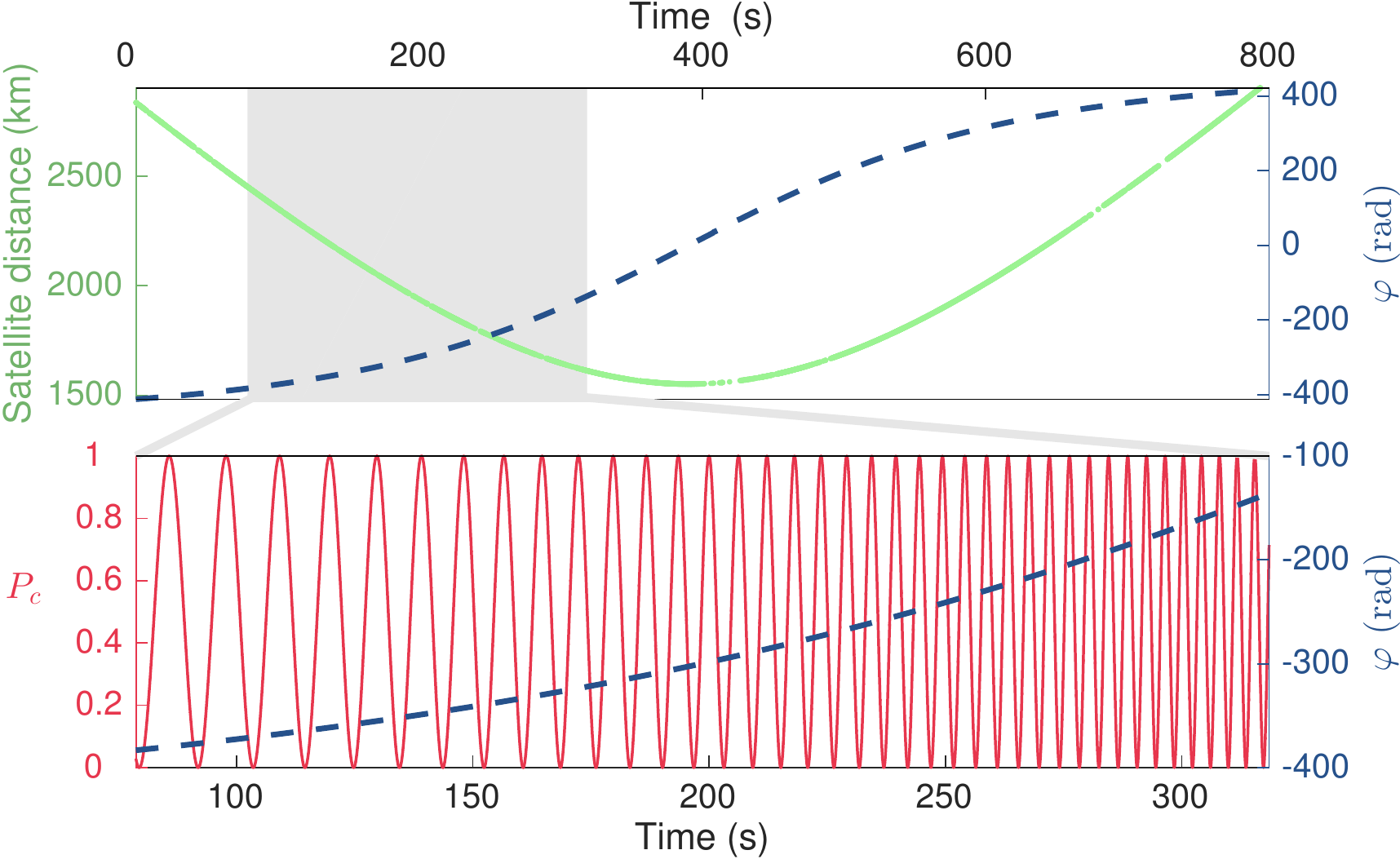}
\caption{
{\bf Kinematic phase and interference pattern}. Top panel: we  show the measured satellite distance and the predicted kinematic phase $\varphi(t)$ 
estimated by eq. \eqref{varphi} as a function of time for a passage of the Ajisai satellite. 
Shaded area represents the temporal window of data acquisition.
Bottom panel:  kinematic phase $\varphi(t)$ and theoretical probability $P_c(t)$ in the shaded area.
The interference pattern is modulated according to the value of $\varphi(t)$ determined by the satellite velocity.}
\label{fig:2}
\end{figure}

As we now explain, the relative phase $\varphi(t)$ is determined by the satellite instantaneous radial velocity with respect to ground, $v_r(t)$. Indeed, at a given instant $t$, the satellite motion determines a shift ${\delta r (t)}$ of the reflector radial position, during the separation $\Delta t$ between the two wavepackets. This shift can be estimated at the first order as 
${\delta r (t)}
\approx v_r(t)\Delta t$, and its value  may reach a few tens of micrometers  for the satellites here used.
For instance, in the top panel of Fig. \ref{fig:1}, we show the value of $v_r(t)$, that ranges from -6 to 6 km/s for the selected passage of the Beacon-C satellite.
Therefore, the satellite motion
imposes during reflection the additional {\it kinematic phase} 
$\varphi(t) \approx 2 \delta r(t)(2\pi/\lambda) $
between the wavepackets $\ket{L}$ and $\ket{S}$, where
 $\lambda$ is the pulse wavelength in vacuum (see Fig.  \ref{fig:2}).

A single MZI for state generation and detection intrinsically ensures the same unbalance
 of the arms and avoids active stabilization, necessary otherwise with two independent interferometers.
As detailed in Appendix \ref{design},
two $4f$-systems realizing an optical relay equal to the arm length difference were
placed in the long arm of the MZI.
The relay is required to match the interfering beam wavefronts that 
are distorted by the passage through atmospheric turbulence: otherwise, the latter 
may cause distinguishability between the two paths,
washing out the interference.
The MZI at the receiver 
is able to reveal the interference between the
two returning wavepackets. 
At the MZI outputs we expect detection times that follow the well known three-peak profile (see Fig. \ref{fig:1}):
the first peak represents the pulse
$\ket{S}$
taking again the short arm, while the third 
represents the delayed pulse $\ket{L}$ taking again the long arm. 
In the central peak we expect indistinguishably between two {\it alternative possibilities}: 
the $\ket{S}$ pulse taking the long arm and the
$\ket{L}$ pulse taking the short arm in the path along the MZI toward the detector.
The signature of interference at the single photon level is then obtained when the counts in the central peak differ
from the sum of the counts registered in the lateral peaks.

To measure the interference we used a single photon detector (PMT) placed 
at the available port of the MZI, as shown in Fig. 1.
For a moving retroreflector, as detailed in Appendix \ref{calculation}, 
special relativistic calculations show that
the probability $P_c$ of detecting the
photon in the central peak  is given by
\beq
\label{Pc_main}
P_c(t)={\frac12}\left[1-\mathcal V(t) \cos\varphi(t)\right]\,,
\eeq
with
\beq
\begin{aligned}
\label{varphi}
\varphi(t)&=\frac{2\beta(t)}{1+\beta(t)}\frac{2\pi c}{\lambda}\Delta t
\\
\mathcal V(t)&=e^{-\,\frac{\lambda^2\varphi^2(t)}{8\pi c^2\tau^2_c}}
=e^{-2\pi\left(\frac{\Delta t}{\tau_c}\frac{\beta(t)}{1+\beta(t)}\right)^2}\simeq 1\, .
\end{aligned}
\eeq

We note that for a retroreflector at rest we expect $P_c=0$. The parameter 
$\beta(t)$ is defined as $\beta(t)=\frac{v_r(t)}{c}$. 
The above relation is 
obtained by time-of-flight calculations together with the Doppler effect that changes the
 angular frequency of the reflected pulses from
 $\omega_0\equiv\frac{2\pi c}{\lambda}$ to $\frac{1-\beta}{1+\beta}\omega_0$.
We note that the
first order approximation of eq. \eqref{varphi}
gives the phase $\varphi(t)\approx4\pi v_r(t)\Delta t/\lambda$ above described.
The theoretical visibility $\mathcal V(t)$ is
approxi\-mately 1 since the $\beta$ factor is upper bounded by $3\cdot10^{-5}$
in all the experimental studied cases, 
while the ratio $\Delta t/\tau_c$
is of the order of $10^2$.

{\it Experimental results - } We realized our  experiment  at the Matera Laser Ranging Observatory (MLRO) of the Italian Space Agency in Matera, Italy,
 that is equipped with a 1.5 m telescope designed for precise satellite tracking and which acted as ground quantum-hub 
 for the first demonstrations of Space 
Quantum Communication (QC)~\cite{vill08njp,vall15prl}. 
  The pulses used to prepare $\PsiOut$
 are generated
by  a mode-locking laser based on a Nd:YVO$_4$ gain medium  operating at a  repetition rate stabilized at $100$ MHz by an atomic clock  and at the wavelength of $1064$ nm.
Each pulse is upconverted  with a PPLN crystal to 
 a wavelength of $532$ nm and energy $\sim 1$ nJ. The pulses, after the MZI, are sent to the Coud\'e path of the MLRO telescope, that directs the state  $\PsiOut$ toward the satellite while actively tracking its orbit. We selected three satellites in low-Earth-orbit (LEO) -- Beacon-C, Stella and Ajisai -- 
which  are equipped with efficient cube-corner retroreflectors (CCR).
Thanks to the CCR properties, the state is automatically redirected toward the ground station,
where it is injected into the same MZI used in the uplink.

The value of $\varphi(t)$ originating from the satellite 
motion can be precisely predicted on the base of the sequence of measurements of the 
 instantaneous distance of the satellite, or $range$ $r$, which is realized in parallel.
 The range is measured by 
a strong Satellite Laser Ranging (SLR) signal at 10 Hz and
energy per pulse of 100 mJ.
Thanks to an atomic clock, the SLR pulses are separated precisely by $\Delta T=100$ ms and synchronized with the 100 MHz pulses.
By measuring the temporal separation $\Delta T'$
of the SLR pulses at the receiver after the
satellite retroreflection, it is possible to
determine the instantaneous satellite velocity relative to the ground station $v_r(t)$. Indeed, since by the Doppler effect
$\Delta T'=\frac{1+\beta}{1-\beta}\Delta T$,
the velocity $v_r(t)$ can be estimated as
$
v_r(t)=c\frac{\Delta T'-\Delta T}{\Delta T'+\Delta T}$.
The separation $\Delta T'$ is related to the
 range $r$ by $\Delta T'=\Delta T+\Delta r/c$, where $\Delta r$ is the variation
of the satellite distance between two subsequent SLR pulses. 
Then, by measuring the range every $100$ ms,
the instantaneous satellite velocity relative to the ground station $v_r(t)$ can be estimated,
from which $\varphi(t)$ can be derived by Eq. \eqref{varphi}.
 In the top panel of Fig. \ref{fig:2}, 
 for a given passage of the  Ajisai satellite, 
 we show the measured satellite distance and the estimated $\varphi(t)$ as a function of  time from the 
 beginning to the end of the satellite tracking.
Since $v_r(t)$ is continuously changing along the orbit,
the value of $\varphi(t)$ is varying accordingly.
In the bottom panel of Fig. 2 we show the variation of the theoretical 
output probability $P_c(t)$
along the Ajisai orbit
as predicted by eq. \eqref{Pc_main}.

\begin{figure}[t]
\centering
\includegraphics[width=8.5cm]{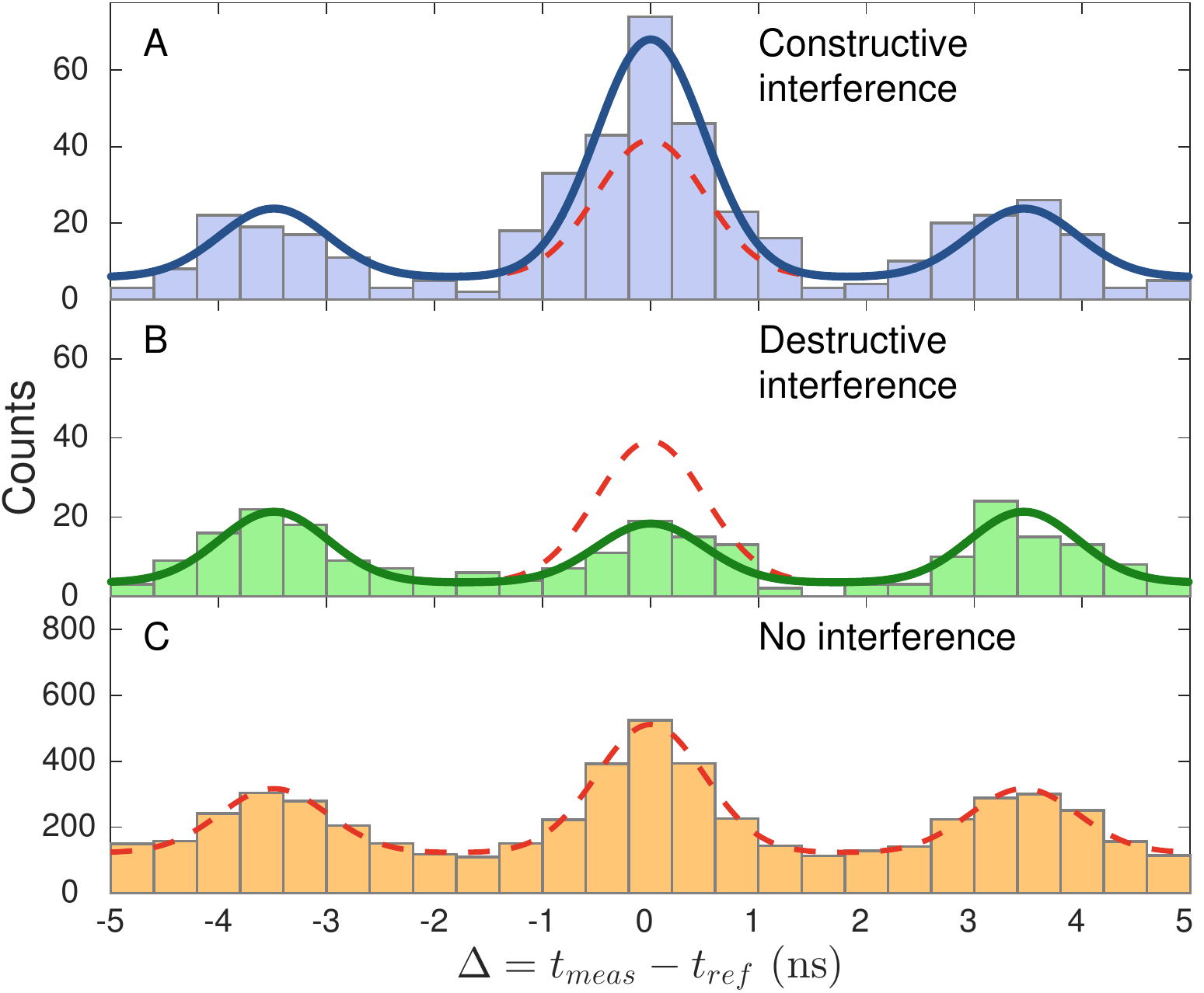}
\caption{{\bf Constructive and destructive {  single photon} interference} (Beacon-C satellite, 11.07.2015 h 1.33 CEST). 
{\bf (A)} Histogram of  single photon detections as a function
of time $\Delta = t_{meas} - t_{ref}$ realized by selecting
only the returns characterized by  $\varphi(\mod2\pi)\in [4\pi/5, 6\pi/5]$
that lead to constructive interference. Solid line shows the tri-Gaussian fit.
By evaluating the Gaussian integrals we
obtained the counts $N_\ell =112\pm11$ for the sum of lateral peaks 
and $N_c = 196\pm14$ for the central one. 
{\bf (B)} Histogram of  single photon detections realized by selecting
only the returns characterized by $\varphi(\mod2\pi) \in [-\pi/5, \pi/5]$.
Here $N_\ell =112\pm11$ and $N_c= 46\pm7$. 
{\bf (C)} Histo\-gram of  single photon detections without any selection on the phase. 
As expected, interference is completed washed out and we measured $N_c=1245\pm35$ and $N_\ell=1306\pm36$,
fully compatible with $P_c=1/2$.
In all panels, dotted red lines represent the expected counts in case of no interference.
}
\label{fig:3}
\end{figure}

By the synchronization technique described in Appendix \ref{synchro}, we 
determined of the expected ($t_{\rm ref}$) and the measured ($t_{\rm meas}$) instant of arrival 
 of each photon. 
In this way, the histogram of the detections in the temporal window of $10$ ns between two consecutive pulses 
as a function of the temporal difference $\Delta = t_{ \rm meas} - t_{\rm ref}$ can be obtained.
In Fig. \ref{fig:3} we show such histograms corresponding
to constructive and destructive interference in the case of satellite Beacon-C.

In particular, for the constructive interference, Fig. \ref{fig:3}A,
we selected the detections corresponding
to
$\varphi$ (mod $2\pi)\in[{4\pi}/{5},{6\pi}/{5}]$.
For the destructive interference, Fig. \ref{fig:3}B,
we selected a kinematic phase 
$\varphi$ (mod $2\pi)\in[-{\pi}/{5},{\pi}/{5}]$.
The detections in the central peak are respectively higher or lower than
the sum of the two lateral peaks in the two
cases.
We note that the peak width is determined by the detector timing jitter
which has standard deviation $\sigma=0.5$ ns.
These two histograms clearly show the
interference effect in the central peak.
On the contrary, Fig. 3C is obtained by taking all the data without any selection on $\varphi$.
In this case, the interference is completely washed out. 
These results show that, in order to prove the interference effect,
it is crucial to correctly predict the kinematic phase $\varphi$ imposed by the satellite motion.

By using the data of Fig. 3, we experimentally  evaluate the probability $P_c^{(\rm exp)}$ 
as the ratio of the detections 
associated the central peak $N_c$ to
twice the sum $N_\ell$ of the detections 
associated to the side peaks, namely
\beq
P_c^{(\rm exp)}=\frac{N_c}{2N_\ell}\,.
\eeq
The values $P_c^{(\rm exp)}=0.87\pm0.10$ and
$P_c^{(\rm exp)}=0.20\pm0.03$
are obtained for constructive and destructive interference respectively. 
The values deviates with clear statistical evidence from 0.5, which is the expected
value in the case of no interference.

\begin{figure}[t]
\centering
\includegraphics[width=8.5cm]{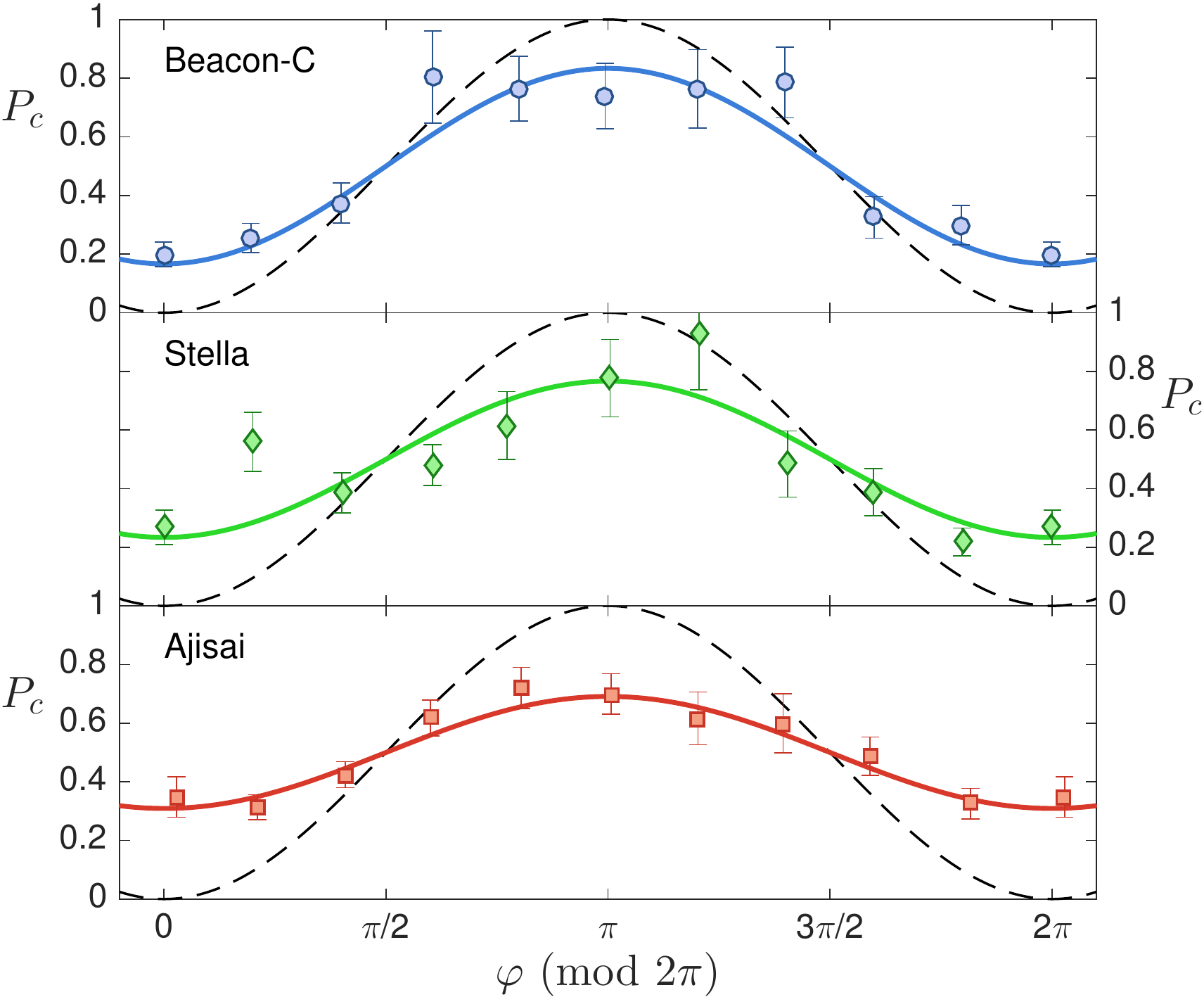}
\caption{ {\bf Experimental interference pattern}.
Experimental probabilities $P^{\rm (exp)}_c$ as a function of the kinematic phase measured for three different satellites.
By fitting the data we
estimate the visibilities
$\mathcal V_{\rm exp}=67\pm11\%$ for Beacon-C,
$\mathcal V_{\rm exp}=53\pm13\%$ for Stella and
$\mathcal V_{\rm exp}=38\pm4\%$ for Ajisai.
Dashed lines correspond to the theoretical value of $P_c$ predicted by eq. \eqref{Pc_main}.
The points are obtained by considering ten intervals of the phase defined by $\mathcal I_j\equiv [\frac{j-1}{10}\pi,\frac{j+1}{10}\pi]$. 
For each interval we selected the data corresponding to
$\varphi(\mod 2\pi)\in\mathcal I_j$:
from such data we
determined the experimental probability of detection in the central peak $P^{\rm (exp)}_c$ and we averaged the corresponding phase $\varphi$. 
We note that at point $\varphi=0$ and $\varphi=2\pi$ the same subset of data were selected.}
\label{fig:4}
\end{figure}

A more clear evidence of the role of $\varphi(t)$ can be demonstrated 
by evaluating the experimental probabilities $P_c^{(\rm exp)}$ as a function of $\varphi$.
 Fig. 4 shows 
$P_c^{(\rm exp)}$ for ten different values of the kinematic phase $\varphi$
and for the three different satellites.
By fitting the data by $P^{\rm (exp)}_c=\frac12(1-\mathcal V_{\rm exp}\cos\varphi)$,
we estimated the experimental visibilities 
$\mathcal V_{\rm exp}=67\pm11\%$ for Beacon-C,
$\mathcal V_{\rm exp}=53\pm13\%$ for Stella and
$\mathcal V_{\rm exp}=38\pm4\%$ for Ajisai.
The data were collected at the following satellite distance ranges:
 from $1600$ to $2500$~km (Ajisai, 12.07.2015, h 3.42 CEST),
 from $1100$ to $1500$~km (Stella, 12.07.2015, h 3.08 CEST) and
 from $1200$ to $1500$~km (Beacon-C, 11.07.2015, h 1.33 CEST),
giving two-way channel lengths ranging from 2200 up to 5000 km.
The interference patterns in  Fig. 4  clearly demonstrate that 
the { coherence between the two temporal modes}
is preserved along these thousand kilometer scale channels with rapidly moving retroreflectors.
 We attribute the different visibilities to residual vibrations of the unbalanced MZI between
the upgoing and downgoing pulses,
since the 
intrinsic visibility of the interferometer was measured to be above $95\%$.
To  improve the visibility it would be necessary to redesign the full interferometric setup to further mitigate this effect.
We note that, in the double-pass configuration, the interferometer is
sensitive to vibrations with frequencies higher than $1/{\rm rtt}$ 
(rtt is the round trip time).
Indeed, 
the lower visibility is obtained with
Ajisai, the satellite with larger distance from the ground (from $1600$ to $2000$ km), with a rtt between 10.7 ms and 16.7 ms. For the other two satellites the rtt
is typically lower than 10 ms. 

The mean number
of photon $\mu$ in the received pulses may be derived
by measuring the detection rate and using the
optical losses $\eta\sim0.27$ in the receiving setup. 
At the primary mirror of the receiving
telescope, the average $\mu$ during the
data acquisition are given by
$\mu\approx7\cdot 10^{-4}$ for Beacon-C,
$\mu\approx2\cdot 10^{-3}$ for Ajisai and
$\mu\approx9\cdot 10^{-4}$ for Stella.
From these values  we may 
conclude that interference was probed at the single photon level.
Indeed, the probability of having more that one photon per pulse
in the receiver MZI is  $\eta\mu^2/2$.
 We estimated that that the mean number of photons $\mu_{\rm sat}$ leaving from the satellites are $\mu_{\rm sat}^{\rm (Stella)}<20$ and $\mu_{\rm sat}^{\rm (Ajisai)}<60$
\cite{beacon}
(the instantaneous values fluctuate due to pointing error and turbulence)
with a total downlink attenuation between 60~dB and 70~dB (including the total detection setup losses $\eta_{rx}\sim 20$~dB 
due to the optical losses $\eta\sim0.27$ and the fact that
we used a single PMT with 10\% efficiency at the output of the MZI).

{\it Conclusions -}
{  Interference at the single photon level} between two temporal modes 
was observed
 along a path that  includes a rapidly moving retroreflector on a satellite
and with length up to 5000 km. 
We have experimentally demonstrated that the relative motion of the satellite with respect to the ground induces a varying phase that modulates
 the interference pattern. This varying phase is not present in the case of fixed terminals. 
The effect resulted from the measured interference pattern during passages of three satellites, Beacon-C, Stella and Ajisai, 
having different relative velocities and distances from MLRO ground station. 

Up to now, photon polarization 
was the only degree of freedom exploited in long distance free-space  Quantum Communications,
along ground links of 143 km~\cite{ursi07nap, ma12nat, capr12prl} 
or in efforts towards quantum key distribution in space channels, of length as
large as 2000 km, as recently demonstrated by our group~\cite{vall15prl}.
Indeed, time-bin was never implemented for QC over a long-distance free-space channels, fearing that turbulence effects on the wavefront may spoil the interference.
Here we have demonstrated that atmospheric turbulence is not detrimental for time-bin encoding in long distance free-space propagation. Indeed, the two temporal modes separated by a few nanoseconds are identically distorted by the propagation in turbulent air, whose dynamics is in the millisecond scale~\cite{capr12prl}: the key point here is the careful matching of the interfering wavefronts in the two arms, as shown in Fig. 1 and 5(A). 
The results here presented attest the feasibility of time-bin/phase encoding technique in the context of Space Quantum Communications.

Furthermore,
the measurement of interference
in Space is a milestone to investigate one of the big unresolved puzzle in Physics,
namely the interplay of Quantum Theory with Gravitation.
As recently proposed theoretically by M. Zych {\it et al.}~\cite{zych11nco,ride12cqg}
(the optical version of the original Colella-Overhauser-Werner (COW) experiment realized with neutrons~\cite{cole75prl}),
 interference with single photons in Space is a witness of general relativistic effects:
gravitational phase shift 
between a superposition of two photon wavepackets
could  be highlighted in the
context of large distance quantum optics experiment. In the case of our setup,
the gravitational shift for the Ajisai satellite corresponds to about 2~mrad (see eq. (23) of~\cite{ride12cqg}).
We point out that, unlike the case of effects manifested by photon polarization rotation,
such small gravitational effects may get better highlighted by using true single photons and by increasing the temporal separation of the two 
interfering modes (with the increase of the difficulty in stabilizing the interferometers with large imbalancement).
To reveal the effect of Gravity in quantum experiments
several other proposals have been presented~\cite{ride12cqg}:
these include the exchange of elementary particles from moving and accelerated reference frames, 
which would allow to test Bell's inequalities~\cite{bell64phy,clau69prl} 
and wavefunction collapse 
and possible gravity-induced decoherence~\cite{ghir90pra} 
in laser interferometry with a long baseline.

The interference patterns measured in the present 
experiment demonstrate that 
{  a coherent superposition between 
two temporal modes} holds in the photon propagation and its interference can be indeed observed over very long channels  involving moving terminals at  fast
relative velocity.
 We believe that the results here described attest the viability of the use of 
 {  temporal modes of light}
for fundamental tests of Physics and
 Quantum Communications around the planet and beyond.

\acknowledgements
We would like to thank Francesco Schiavone, Giuseppe Nicoletti, and the MRLO technical operators for the collaboration and support,  Prof.~Roberto Regazzoni of INAF-Osservatorio Astronomico di Padova for the useful discussion on the interferometer optics as well as Dr. Davide Bacco and Simone Gaiarin for their contributions to the setup. We also thank Franco Ambrico for the image  of MLRO.
Our work was supported by the Strategic-Research-Project QUINTET of the Department of Information Engineering, University of Padova, the Strategic-Research-Project QUANTUMFUTURE of the University of Padova.

{\it Note added - }
After the completion of our work, a
laboratory experiment 
 for exploiting
time-bin encoding after free-space propagation,
has been reported~\cite{jin15qph}.

\appendix
\clearpage
\onecolumngrid

\setcounter{equation}{0}
\renewcommand{\theequation}{A\arabic{equation}}

\section{Experimental design}
\label{design}
\begin{figure}[h]
\centering\includegraphics[width=14cm]{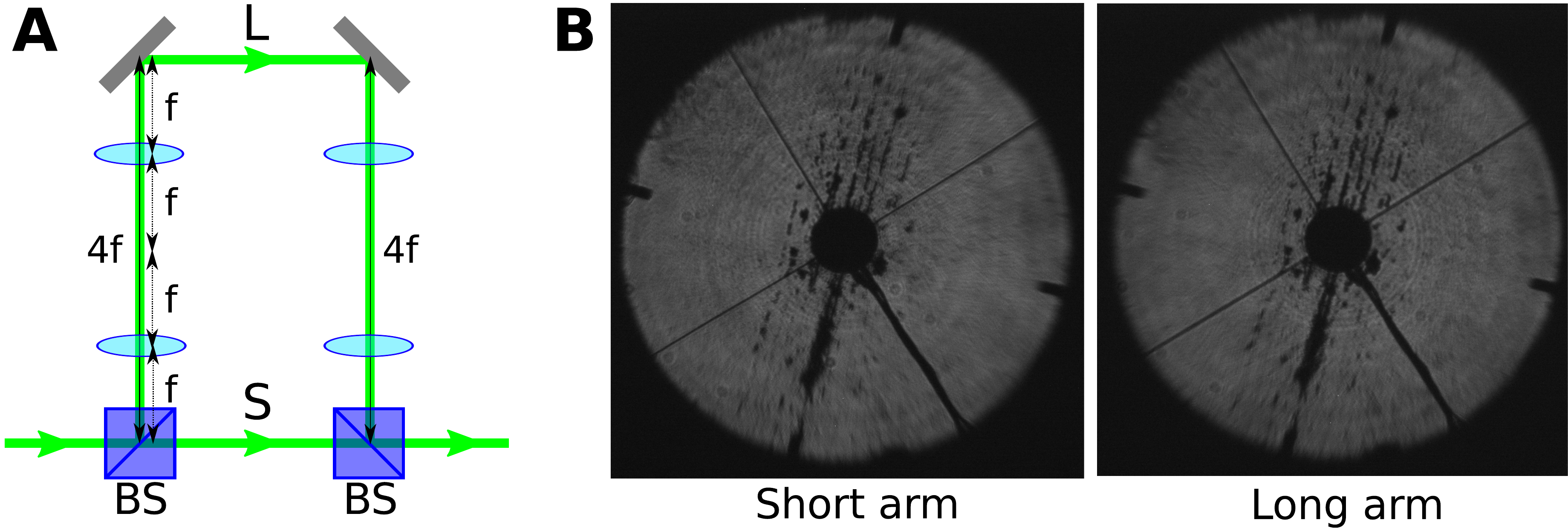}
\caption{{\bf (A)} Scheme of the unbalanced Mach-Zehnder interferometer used in the experiment.
{\bf (B)} Images of the primary mirror with  only the  short or long arm opened.
{We exploited the secondary mirror spider and the condensation spots on the primary mirror surface
for checking the alignment of the interferometer.} }
\label{4fsystem}
\end{figure}

We give a detailed description of the unbalanced Mach-Zehnder interferometer used for the generation and measurement of the
interference. The interferometer,  realized by two beam splitters (BSs), is  schematically shown in Fig. \ref{4fsystem}A.
The difference between the short (S) and long (L) arm is approximately 1 m. 

In order to observe interference  it is necessary to perfectly match the wavefronts of the wave packets traveling 
through the short or the long arm. 
This matching is particularly necessary for the photon beams reflected by the satellite that are subjected to the atmosphere distortion.
To this purpose, we exploited two $4f$-systems that realize an optical relay of length equal to the path difference between 
the long and short arm  of the MZI. Each $4f$-system is composed by two lenses with focal length $f=125$ cm, as
shown in Fig. \ref{4fsystem}A. It is worth noticing that a single $4f$-system, while realizing an optical relay of length $4f$, realizes also
a mirror transformation on the wavefront. A second $4f$-system is thus necessary to compensate this mirror transformation.

In Fig. \ref{4fsystem}B we show the performance of the two $4f$-systems. At the output of the MZI we imaged
the primary mirror while the telescope was pointing at a bright star. The two images in Fig. \ref{4fsystem}B are obtained by blocking the long and short 
arm respectively. The two images are comparable, showing that the wavefronts traveling the short or long arm are well matched
at the output of the interferometer.

\section{Temporal synchronization}
\label{synchro}
The quantum measurement at the receiver requires a very precise temporal synchronization and a strong rejection of the background, necessary to observe the single photon interference. The measurement of the satellite position 
along the orbit by a laser ranging  technique
as in Fig. 2 of the main text was also essential for the determination of the expected instant of arrival 
($t_{\rm ref}$) of the photons
at the MZI output. 
By using a time-to-digital converter with $81$ ps resolution (QuTAU), we acquired the start and stop signals of the  laser ranging  pulses, 
together with PMT detections. We denote PMT timestamps as $t_{\rm meas}$. In this way, 
we may calculate the histogram of the returns in the temporal window of $10$ ns between two consecutive pulses 
as a function of the temporal difference $\Delta = t_{ \rm meas} - t_{\rm ref}$ 
corresponding to a desired value of $\varphi(t)$.

\section{Interference by a mirror moving at constant velocity.}
\label{calculation}
Here we report the detailed calculations used to obtain equations (1) and (2) of the main text. 
Indeed, we will derive the effect of the reflection by a mirror moving at constant velocity on the superposition between two photon  temporal modes.

Let's consider a single photon wavepacket  whose peak
passes at $r=0$ for $t=0$. 
Its wavefunction is given by
\beq 
\psi_0(\tau_-)=\sqrt[4]{\frac{2}{\tau_c^2}}e^{-\pi\frac{\tau_-^2}{\tau^2_c}}e^{i\omega_0\tau_-}\,,
\eeq 
where we have defined the quantities
\beq 
\tau_\pm=\frac{r}{c}\pm t\,,
\eeq 
and $\omega_0=\frac{2\pi c}{\lambda}$ is the angular frequency.
In our convention $r$ is the direction of propagation.
The parameter $\tau_c$ represents the coherence time of the pulse.
We have used the convention of \cite{Saleh-Teich}, namely
\beq
\tau_c=\int|g(\tau)|^2\dd \tau\,,
\eeq 
\beq 
g(\tau)=\langle\psi^*(t)\psi(t+\tau)\rangle=
\int^{+\infty}_{-\infty} \psi^*_0(t)\psi_0(t+\tau)\dd t=
e^{-\frac{\pi\tau^2}{2\tau_c^2}}e^{i\omega_0\tau}\,.
\eeq

If the pulse passes through the unbalanced Mach-Zehnder interferometer shown in Fig. \ref{4fsystem}A, the output is given by
\beq
\psi_1(\tau_-)=\frac{1}{\sqrt{2}}[\psi_0(\tau_-)-\psi_0(\tau_--\Delta t)]
=\frac{1}{\sqrt[4]{2\tau^2_c}}\left[e^{-\pi\frac{\tau_-^2}{\tau_c^2}}-e^{-\pi\frac{(\tau_--\Delta t)^2}{\tau_c^2}}e^{-i\omega_0\Delta t}\right]
e^{i\omega_0\tau-}\,,
\label{psi1} \eeq 
where $\Delta t$ is the MZI unbalancement.

\begin{figure}[h]
\centering\includegraphics[width=12cm]{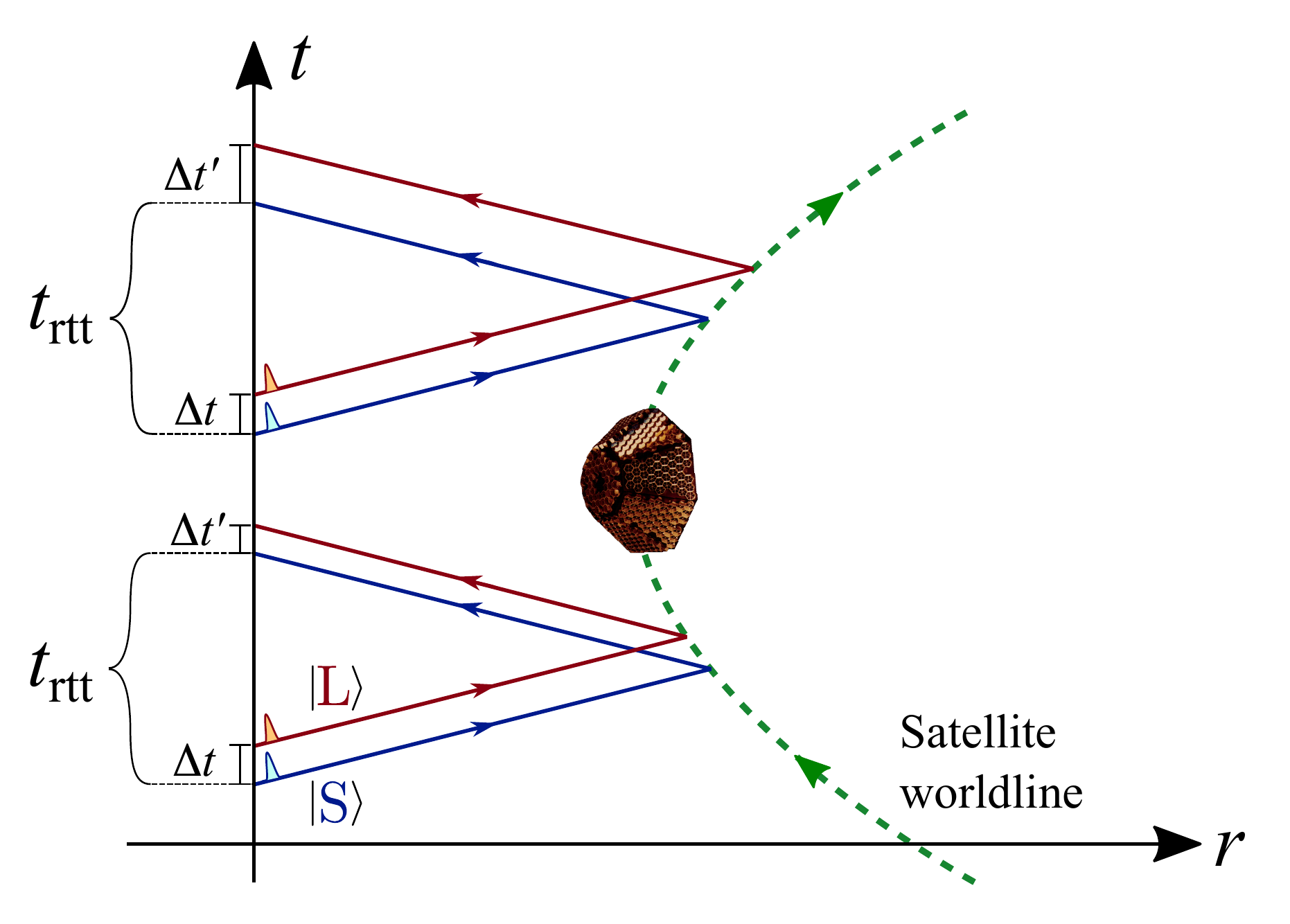}
\caption{{\bf Spacetimes diagrams of light propagation}. Short and Long arm pulses are represented by blue and red color respectively.
Dashed green line represents the satellite trajectory. As explained in the text, pulses separated on ground by a delay $\Delta t$ are
received with a delay $\Delta t'=\Delta t/f_\beta$ due to the motion of the satellite. The round trip time (rtt) is $t_{\rm rtt}$.
Satellite distance $r$ and time $t$ are not to scale.}
\label{spacetime}
\end{figure}

We now consider a mirror on a satellite moving at constant velocity with respect to the interferometer that at time $t=0$ is located
at $r=r_{\rm sat}$. We may change reference frame by setting the origin at the location of the satellite.
The corresponding Lorentz transformation are given by
$$
\left\{
\begin{aligned}
r'&=\gamma(r-r_{\rm sat}-\beta ct)
\\
t'&=\gamma(t-\beta\frac{r-r_{\rm sat}}{c})
\end{aligned}
\right.
\qquad\,,\qquad
\left\{
\begin{aligned}
r&=r_{\rm sat}+\gamma(r'+\beta ct')
\\
t&=\gamma(t'+\beta\frac{r'}{c})
\end{aligned}
\right.
$$
and where $\beta$ has been defined after Eq. 3 and $\gamma = (1 - \beta^2)^{-\frac{1}{2}}$ is the Lorentz factor.
In the upward path, the state $\psi_1(\tau_-)$ is directed by the telescope toward the mirror and
the wave function in the mirror reference frame can be derived by the transformations of the $\tau_\pm$ parameters:
\beq
\label{tau}
\tau_\pm=\gamma(1\pm\beta)\tau'_\pm+\frac{r_{\rm sat}}{c}=\sqrt{\frac{1\pm\beta}{1\mp\beta}}\tau'_\pm+\frac{r_{\rm sat}}{c}\,.
\eeq
In the mirror
reference frame,  the mirror reflection can be  simply described by the transformation $\tau'_-\rightarrow-\tau'_+$.
We now use \eqref{tau}, namely $\tau'_+=\frac{1}{\gamma(1+\beta)}(\tau_+-\frac{r_{\rm sat}}{c})$ to go back to the interferometer reference frame.
If we define $t_{\rm rtt}=\frac{2}{1-\beta}\frac{r_{\rm sat}}{c}$,
the total transformation can be summarized by
$$
\begin{aligned}
\tau_-
&\xrightarrow{\quad {\rm boost\ to\ mirror\ ref.\ frame}\quad}\gamma(1-\beta)\tau'_-+\frac{r_{\rm sat}}{c}
\xrightarrow{\quad {\rm reflection}\quad} -\gamma(1-\beta)\tau'_++\frac{r_{\rm sat}}{c}
\\
&\xrightarrow{\quad {\rm boost\ back\ to\ ground\ ref.\ frame}\quad}-f_\beta(\tau_+-t_{\rm rtt})\,,
\\
\end{aligned}
$$
where
$$f_\beta=\gamma^2(1-\beta)^2=\frac{1-\beta}{1+\beta}\,.$$
It is worth noticing that also normalization should be changed in order to preserve normalization.
The beam coming back from the satellite is then written as
\beq
\begin{aligned}
\psi_2(\tau_+)&=\sqrt{f_\beta}\psi_1(-f_\beta(\tau_+-t_{\rm rtt}))
\\&=
\frac{\gamma(1-\beta)}{\sqrt{2}}[\psi_0(-f_\beta(\tau_+-t_{\rm rtt}))
-\psi_0( -f_\beta(\tau_+-t_{\rm rtt})-\Delta t)]\,.
\end{aligned}
\label{back_pulses}
\eeq
We now comment the two terms. 
The first term is
\beq
\label{first_pulse}
\psi_0(-f_\beta(\tau_+-t_{\rm rtt}))=
\sqrt[4]{\frac{2}{\tau_c^2}}e^{-\pi\frac{f^2_\beta(\tau_+-t_{\rm rtt})^2}{\tau_c^2}}\,
e^{-if_\beta\omega_0(\tau_+-t_{\rm rtt})}\,,
\eeq
representing a pulse with a coherence time stretched (or compressed) by the Doppler effect: its coherence time is indeed
\beq
\tau'_c=\frac{\tau_c}{f_\beta}=\frac{1+\beta}{1-\beta}\tau_c\,.
\eeq
We also note that the parameter $t_{\rm rtt}$ represents the time that the pulse peak takes to come back to the origin (namely the round trip time).
The second term in \eqref{back_pulses} represents the pulse of eq. \eqref{first_pulse} 
delayed by $\Delta t'=\frac{\Delta t}{f_\beta}$. 
The same relation applies also to the SLR
pulses, that are separated in time by $\Delta T=100$ ms. 
As detailed in the main text,
by measuring the temporal separation $\Delta T'$
of the SLR pulses at the receiver, it is possible to
determine the satellite velocity.
Spacetime diagrams for the 
pulses going upward to and downward from the satellite are shown in Fig. \ref{spacetime} for different parts of the satellite orbit.

After passing again in the interferometer we get at the detection port of the MZI (see Fig. 1 of the main text) the following state:
$$
\begin{aligned}
\psi_3(\tau_++t_{\rm rtt})=
\frac{i\gamma(1-\beta)}{2}&\Big[\psi_0(-f_\beta \tau_+)+
\psi_0(-f_\beta(\tau_++\Delta t))
\\
&\quad-\psi_0(- \Delta t-f_\beta\tau_+)
-\psi_0(-\Delta t-f_\beta(\tau_++\Delta t))
\Big]\,.
\end{aligned}
$$
We used the convention that a BS introduces a $i$ phase shift for the reflected beam and does not change the transmitted beam.

We now have three pulses at the detector: the probability of getting the photon in the central pulse at $r=0$  is given by
\beq
\begin{aligned}
P_c(t)&=\frac{\gamma^2(1-\beta(t))^2}{4}\int\dd t'|\psi_0(-f_\beta(t'+\Delta t))-\psi_0(-\Delta t-f_\beta t')|^2
\\
&=\frac12\left\{1-\sqrt{\frac{2}{\tau_c^2}}\int\dd t' 
\Re e\left[
e^{-\pi\frac{(t'+f_\beta\Delta t)^2}{\tau_c^2}}
e^{-\pi\frac{(t'+\Delta t)^2}{\tau_c^2}}
e^{i\omega_0(1-f_\beta)\Delta t}\right]\right\}
\\
&={\frac12}\left[1-\mathcal V(t) \cos\varphi(t)\right]\,,
\end{aligned}
\eeq
with
\beq
\varphi(t)=\omega_0[1-f_{\beta})]\Delta t
=\frac{2\beta(t)}{1+\beta(t)}\omega_0\Delta t\,,
\eeq
and
\beq
\begin{aligned}
\mathcal V(t)&=
\sqrt{\frac{2}{\tau_c^2}}
\int\dd t'\,
e^{-\pi\frac{(t'+f_\beta\Delta t))^2}{\tau_c^2}}
e^{-\pi\frac{(t'+\Delta t)^2}{\tau_c^2}}
=\exp\{-2\pi\left[\frac{\Delta t}{\tau_c}\frac{\beta(t)}{1+\beta(t)}\right]^2\}\,.
\end{aligned}
\eeq
The above results give  equations (1-2) of the main text.


\begin{thebibliography}{10}
\bibitem{beacon}
{  The technique used for the estimation of the $\mu_{\rm sat}$ is not applicable
for the Beacon-C satellite, due to the lack of data (it was launched on 1965). We note that its distance to the ground is similar to the Stella satellite and a similar 
$\mu_{\rm sat}$ could be expected.}
\end{thebibliography}
\end{document}